\documentclass[twocolumn]{aastex62}
\usepackage[normalem]{ulem}
\useunder{\uline}{\ul}{}
\usepackage{hyperref}
\usepackage{lineno}

\usepackage{color}
\usepackage{amsmath,amssymb}

\definecolor{gray1}{gray}{0.8}
\definecolor{gray2}{gray}{0.8}
\definecolor{gray3}{gray}{0.8}

\newcommand{\pone}{Paper--I}

\hypersetup{linkcolor=cyan,citecolor=magenta,filecolor=yellow,urlcolor=blue}

\revised{\today}

\shorttitle{}
\shortauthors{}

\begin{document}

%%%%%%%%%%%%%%%%%%%%%%%%%%%%%%%%%%%%%%%%%%%%%%%%%%

%%%%%%%%%%%%%%%%%%%%%%%%%%%%%%%%%%%%%%%%%%%%%%%%%%

%%%%%%%%%%%%%%%%%%% TITLE PAGE %%%%%%%%%%%%%%%%%%%

% Title of the paper, and the short title which is used in the headers.
% Keep the title short and informative.
\title[]{On the X-ray, optical and radio afterglows of the BdHN I GRB 180720B generated by the synchrotron emission}
%On the X-ray, optical and radio afterglow of GRB 180720B generated by the synchrotron emission of the BdHN progenitor

\author{J.~A.~Rueda}
\affiliation{ICRANet, Piazza della Repubblica 10, I-65122 Pescara, Italy}
\affiliation{ICRA, Dip. di Fisica, Sapienza Universit\`a  di Roma, Piazzale Aldo Moro 5, I-00185 Roma, Italy}
\affiliation{ICRANet-Ferrara, Dip. di Fisica e Scienze della Terra, Universit\`a degli Studi di Ferrara, Via Saragat 1, I--44122 Ferrara, Italy}
\affiliation{Dip. di Fisica e Scienze della Terra, Universit\`a degli Studi di Ferrara, Via Saragat 1, I--44122 Ferrara, Italy}
\affiliation{INAF, Istituto di Astrofisica e Planetologia Spaziali, Via Fosso del Cavaliere 100, 00133 Rome, Italy}

\author{Liang~Li}
\affiliation{ICRANet, Piazza della Repubblica 10, I-65122 Pescara, Italy}
\affiliation{ICRA, Dip. di Fisica, Sapienza Universit\`a  di Roma, Piazzale Aldo Moro 5, I-00185 Roma, Italy}
\affiliation{INAF, Osservatorio Astronomico d'Abruzzo, Via M. Maggini snc, I--64100, Teramo, Italy}

\author{R.~Moradi}
\affiliation{ICRANet, Piazza della Repubblica 10, I-65122 Pescara, Italy}
\affiliation{ICRA, Dip. di Fisica, Sapienza Universit\`a  di Roma, Piazzale Aldo Moro 5, I-00185 Roma, Italy}
\affiliation{INAF, Osservatorio Astronomico d'Abruzzo, Via M. Maggini snc, I--64100, Teramo, Italy}

\author{R.~Ruffini}
\affiliation{ICRANet, Piazza della Repubblica 10, I-65122 Pescara, Italy}
\affiliation{ICRA, Dip. di Fisica, Sapienza Universit\`a  di Roma, Piazzale Aldo Moro 5, I-00185 Roma, Italy}
\affiliation{Universit\'e de Nice Sophia-Antipolis, Grand Ch\^ateau Parc Valrose, Nice, CEDEX 2, France}
\affiliation{INAF,Viale del Parco Mellini 84, 00136 Rome, Italy}

\author{N.~Sahakyan}
\affiliation{ICRANet, Piazza della Repubblica 10, I-65122 Pescara, Italy}
\affiliation{ICRANet-Armenia, Marshall Baghramian Avenue 24a, Yerevan 0019, Republic of Armenia}

\author{Y.~Wang}
\affiliation{ICRANet, Piazza della Repubblica 10, I-65122 Pescara, Italy}
\affiliation{ICRA, Dip. di Fisica, Sapienza Universit\`a  di Roma, Piazzale Aldo Moro 5, I-00185 Roma, Italy}
\affiliation{INAF, Osservatorio Astronomico d'Abruzzo, Via M. Maggini snc, I--64100, Teramo, Italy}

\email{jorge.rueda@icra.it, liang.li@icranet,org, rahim.moradi@icranet.org\\ ruffini@icra.it, narek.sahakyan@icranet.org, yu.wang@icranet.org}

\begin{abstract}

Gamma-ray bursts (GRBs) are systems of unprecedented complexity across all the electromagnetic spectrum, including the radio, optical, X-rays, gamma-rays in the megaelectronvolt (MeV) and gigaelectronvolt (GeV) regime, as well as ultrahigh-energy cosmic rays (UHECRs), each manifested in seven specific physical processes with widely different characteristic evolution timescales ranging from $10^{-14}$~s to $10^{7}$~s or longer. We here study the long GRB 180720B originating from a binary system composed of a massive {carbon-oxygen} (CO) star of about $10 M_\odot$ and a companion neutron star (NS). The gravitational collapse of the CO star gives rise to a spinning newborn NS ($\nu$NS), with an initial period of $P_0=1$~ms that powers the synchrotron radiation in the radio, optical, and X-ray wavelengths. We here investigate solely the GRB 180720B afterglows and present a detailed treatment of its origin based on the synchrotron radiation released by the interaction of the $\nu$NS and the SN ejecta. We show that in parallel to the X-ray afterglow, the spinning $\nu$NS also powers the optical and radio afterglows and allows to infer the $\nu$NS and ejecta parameters that fit the observational data. %The core concepts of the underlying synchrotron radiation are presented.
\end{abstract}



%%%%%%%%%%%%%%%%%%%%%%%%%%%%%%%%%%%%%%%%%%%%%%%%%%
%%%%%%%%%%%%%%%%% BODY OF PAPER %%%%%%%%%%%%%%%%%%

%%%%%%%%%%%%%%%%%%%%%%%%%%%%%%%%%%%%%%%%%%%%%%%%%%%%
%%%%%%%%%%%%%%%%%%%%%%%%%%%%%%%%%%%%%%%%%%%%%%%%%%%%
\section{Introduction}\label{sec:1}
%%%%%%%%%%%%%%%%%%%%%%%%%%%%%%%%%%%%%%%%%%%%%%%%%%%%
%%%%%%%%%%%%%%%%%%%%%%%%%%%%%%%%%%%%%%%%%%%%%%%%%%%%

{GRB 180720B was observed in the gamma-rays by Fermi-GBM \citep{2018GCN.22981....1R}, CALET Gamma-ray Burst Monitor \citep{2018GCN.23042....1C}, Swift-BAT \citep{2018GCN.22973....1S}, Fermi-LAT \citep{2018GCN.22980....1B} and Konus-Wind \citep{2018GCN.23011....1F}. The High Energy Stereoscopic System (H.E.S.S.) observed the source in the sub-TeV energy domain ($100$--$440$~GeV; \citealp{2019Natur.575..464A}). In the X-rays, the XRT on board the Neil Gehrels Swift Observatory (for short, Swift) observed the source from $91$ s after the Fermi-GBM trigger \citep{2018GCN.22973....1S}, MAXI/GSC started at $296$ s \citep{2018GCN.22993....1N} and NuStar from $243$ ks to $318$ ks \citep{2018GCN.23041....1B}. The $1.5$-m Kanata telescope observed the source in the optical and near infrared as early as $78$ s from the GRB trigger time \citep{2018GCN.22977....1S}. Additional observations in the optical, infrared and radio telescopes can be found in \citet{2018GCN.22976....1M,2018GCN.22977....1S,2018GCN.22983....1I,2018GCN.22985....1K,2018GCN.22988....1C,2018GCN.23017....1W,2018GCN.23020....1S,2018GCN.23021....1C,2018GCN.23023....1L,2018GCN.23024....1J,2018GCN.23033....1Z,2018GCN.23037....1S,2018GCN.23040....1I,2019Natur.575..464A}. The identification of the Fe II and Ni II lines in the optical observations by the VLT/X-shooter telescope led to the source cosmological redshift of $z = 0.654$ \citep{2018GCN.22996....1V}. With the knowledge of the redshift, the GRB 180720B isotropic energy released is $E_{\rm iso}=5.92 \times 10^{53}$~erg \citep{2018GCN.23019....1R, 2019Natur.575..464A, 2019ApJ...885...29F}. For additional details on the GRB 180720B data, we refer the reader to section \ref{sec:3}.
}

{
Therefore, GRB 180720B is specially relevant to test GRB models given the statistical significance of the available multiwavelength observational data. In this article, we analyze the source X-optical-radio afterglow emission. The description of the afterglow emission of GRB 180720B within the traditional jetted fireball model driven by the interaction of internal and external shocks with the surrounding and interstellar medium, we refer, e.g., to \citet{
2019Natur.575..464A, 2019ApJ...885...29F, 2020A&A...636A..55R, 2020MNRAS.496.3326R}. We here focus on the description of the GRB 180720B afterglow emission within the binary-driven hypernova (BdHN) model. 
}

The BdHN scenario proposes that the GRB progenitor is a binary consisting of a carbon-oxygen {(CO)} star and a neutron star (NS) companion of $\sim 2 M_\odot$. The gravitational collapse of the CO star, unlike considerations {of the \textit{collapsar} scenario that} purports the formation of a single black hole (BH) of $\sim 5$--$10~M_\odot$ \citep{1993ApJ...405..273W}, creates in the supernova (SN) explosion a newborn NS ($\nu$NS) of $\sim 1.5~M_\odot$. {For compact CO-NS binaries with orbital periods of the order of a few minutes, the accretion of the SN ejecta onto the companion NS leads to the formation of a black hole (BH) \citep[see, e.g.,][]{2014ApJ...793L..36F, 2015PhRvL.115w1102F}. These systems, called BdHN I, explain the long GRBs with energy release $\gtrsim 10^{52}$ erg. For wider binaries, the accretion is not sufficient for the NS to reach the critical mass for gravitational collapse, so it remains stable as a massive NS. These systems, called BdHN II, are characterized by energies $\lesssim 10^{52}$ erg. In agreement with the above arguments, GRB 180720B has been classified as a BdHN I \citep[see][hereafter Paper--I]{2021arXiv210309158M}. We give further observational and theoretical details below and in section \ref{sec:2}.}

The interaction of the SN ejecta with the $\nu$NS and the companion BH leads to complementary explosive episodes, which have been gradually identified in the last twenty years \citep{2021MNRAS.504.5301R}. {In section \ref{sec:2}, we present a detailed description of the physical processes that occur in BdHNe I and their related observables, based on the ongoing interplay between theoretical developments and numerical simulations to fit up-to-date observations.} In addition to the X-rays, GeV and sub-TeV afterglows, the presence of the six episodes in GRB 180720B, marked by their specific spectra representing different underlying physical processes has been confirmed in \pone {(see Sec. \ref{sec:2} for additional details)}:
\begin{figure*}
\centering
\includegraphics[width=\hsize,clip]{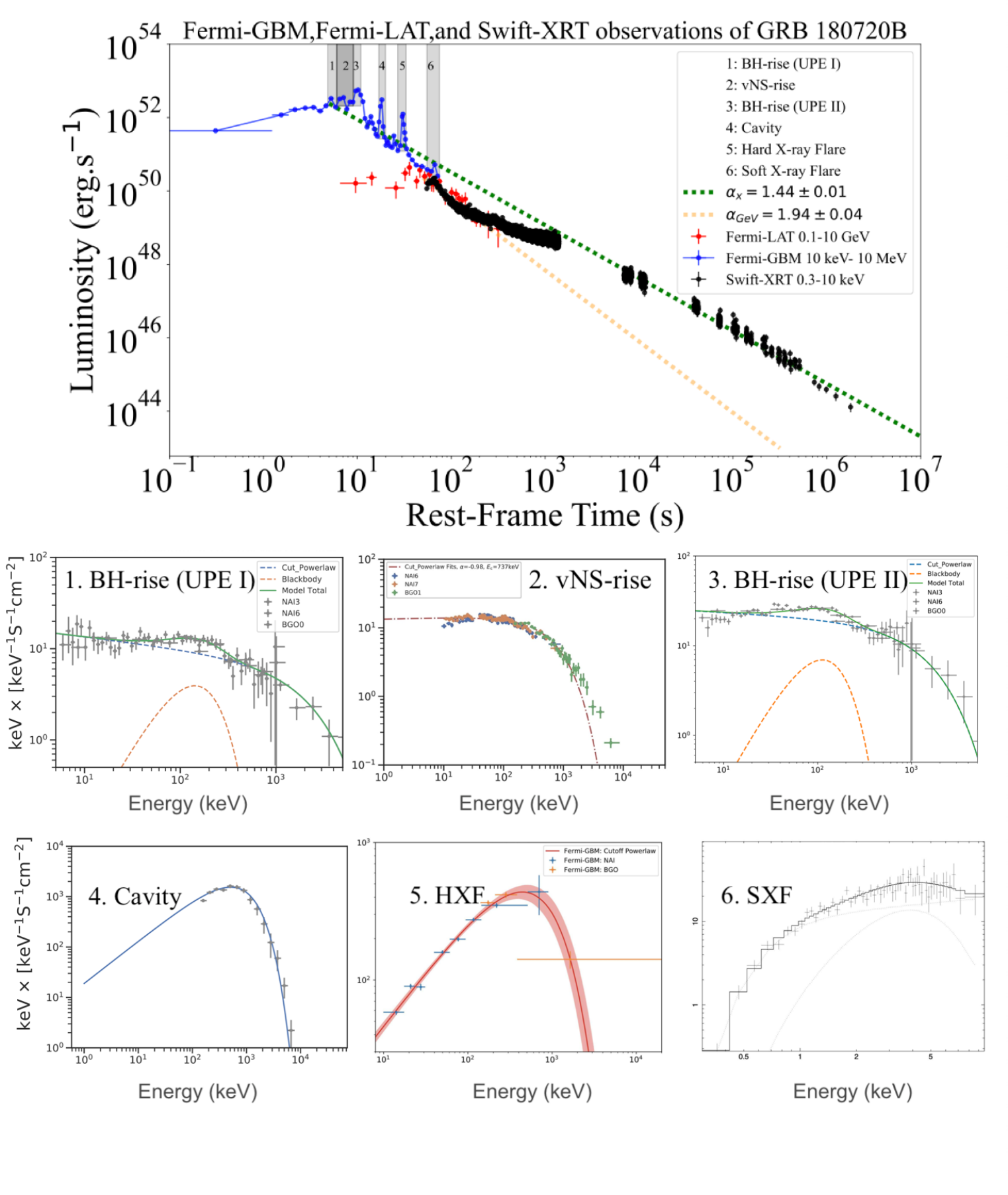}
\caption{Luminosity light-curve of GRB 180720B and spectra related to the different Episodes identified in GRB 180720B. Plots and best fits are reproduced from \pone~by the author's permission. }
\label{fig:lightcurve+spectra}
\end{figure*}

I) Episode 1 (UPE I): the BH formation and its consequent ultra-relativistic emission (UPE) phase originated from the vacuum polarization with its characteristic Lorentz factor $\Gamma \sim 100$ \citep{RSWX,RSWX2}. This marks the first manifestation of the BH (BH-rise). In GRB 180720B, the UPE I extends from $~t_{\rm rf}=4.84$~s to $~t_{\rm rf}=6.05$~s, its isotropic energy is $E^{\rm MeV}_{\rm UPE I}=(6.37\pm0.48) \times 10^{52}$~erg, and its spectrum is best fitted by a CPL+BB model (index $\alpha=-1.13$, cutoff energy $E_{\rm c}=2220.569$~keV, and blackbody (BB) temperature $k T = 50.31$~keV in the observer's frame); see Fig 5A in \pone. 

II) Episode 2 ($\nu$NS-rise): the accretion of ejecta onto the $\nu$NS leads to the $\nu$NS spin-up, which we describe in this article. The first appearance of this interaction in GRB 180720B, referred as the \textit{$\nu$NS-rise}, extends from $~t_{\rm rf}=6,05$~s to $~t_{\rm rf}=9.07$~s, has an isotropic energy of $E^{\rm MeV}_{\rm \nu Ns}=(1.13\pm0.04) \times 10^{53}$~erg, and its spectrum is best fitted by a CPL model ($\alpha=-0.98$, and  $E_{\rm c}=737$~keV, in the observer's frame); see Fig.~5B in \pone.

III) Episode 3 (UPE II): is marked by the observation of the first significant GeV photon at $~t_{\rm rf}=7.06$~s, and includes as well the continuation of the UPE phase from $~t_{\rm rf}=9.07$~s to $~t_{\rm rf}=10.89$~s, with an isotropic energy of $E_{\rm UPE II}^{\rm MeV}=(1.6 \pm 0.95) \times 10^{53}$~erg. The spectrum of the UPE II is best fitted by a CPL+BB model with model parameters of $\alpha= -1.06^{+0.01}_{-0.01}$, $E_{\rm c}=1502.5^{+88.6}_{-87.5}$~keV and $kT= 39.8^{+1.6}_{-1.6}$~keV; see Fig.~5C in \pone.  

IV) Episode 4 (Cavity): this radiation occurs in the cavity formed during the gravitational collapse of the NS and the consequent BH formation, which becomes further depleted by the UPE phase \citep{2019ApJ...883..191R}. For GRB 180720B, it occurs from $t_{\rm rf}=16.94$~s to $~t_{\rm rf}=19.96$~s, with an isotropic energy of $E_{\rm CV}^{\rm MeV}=(4.32 \pm 0.19) \times 10^{52}$~erg, characterized by a CPL spectrum ($\alpha=-1.16$, $E_{\rm c} = 607.96$~keV). Its spectrum is given in Fig.~5D in \pone.

V) Episode 5 (SXF), and VI) Episode 6 (HXF): the interaction of the UPE with the SN ejecta, which has a typical torus-like morphology with a cone of half-opening angle $\sim 60 ^{\circ}$ from the normal to the orbital plane \cite{2021MNRAS.504.5301R}, creates the further emission of hard X-ray flare (HXF) and soft X-ray flare (SXF) \citep{2018ApJ...869..151R}. The HXF of GRB 180720B occurs from $t_{\rm rf}= 28.95$~s to $t_{\rm rf}= 34.98$~s, with $L_{\rm HXF,iso}^{\rm MeV}=(7.8 \pm 0.07) \times 10^{51}$~erg~s$^{-1}$, and is best fitted by a CPL model with $E_{\rm c}=(5.5_{-0.7}^{+0.8}) \times 10^2$~keV, $\alpha = -1.198 \pm 0.031$; see Fig.~5E in \pone. The SXF occurs from $t_{\rm rf}= 55$~s to $t_{\rm rf}= 75$~s, with $L_{\rm SXF,iso}^{\rm X}=1.45\times 10^{50}$~erg, and is best fitted by a PL+BB spectrum with $\alpha = -1.79 \pm 0.23$, and $k T=0.99 \pm 0.13$~keV; see Fig.~5F in \pone.

In GRB {180720B}, as indicated in \pone, the emission processes originate from three independent energy sources, following the evolution of the progenitors composed of a CO$_{\rm core}$ and companion NS:
\begin{enumerate}
    \item The SN. As all SNe, it generates a pulsar which here indicated as $\nu$NS.
    \item The $\nu$NS. The first appearance of the $\nu$NS ($\nu$NS-rise) shows that it has been spun up by fallback accretion of SN ejecta, and is then followed by the late X-ray afterglow with a power-law luminosity $L_X \propto t^{-1.44\pm 0.01}$. From the energetics of the $\nu$NS-rise and the X-ray afterglow, we infer that the $\nu$NS period and magnetic field are $P_{\rm \nu NS,0}=1.01$ ms and $B_{\rm dip}= 4.2 \times 10^{13}$~G.
    \item The BH. The appearance of the BH (BH-rise), formed from the NS companion collapse by accretion of SN ejecta, is marked by the UPE phase followed by the observation of GeV, cavity, HXF and SXF. From the MeV-GeV emission powered by the BH extractable energy, we infer the BH mass $M = 2.4 M_{\odot}$, and spin parameter, $\alpha = 0.55$ (see \pone).
\end{enumerate}
 
During the UPE phase of GRB 180720B, the \textit{inner engine} of the GRB operates in an overcritical regime. The inner engine consists of a Kerr BH of mass $M$ and angular momentum $J$, surrounded by a very low-density plasma of ions and electrons with $\rho_{\rm }\sim 10^{-14}$ g cm$^{-3}$ \citep{2019ApJ...886...82R} in the presence of a uniform magnetic field around the BH, amplified by the gravitational collapse \citep{2020ApJ...893..148R}. Following the quantum vacuum polarization process in decreasing time bins, we have determined the timescale of the radiation ($\tau \sim 10^{-9}$~s), the Lorentz factors ($\Gamma \sim 30$), the radius of transparency ($R^{\rm tr}\sim 10^9$~cm), and energy ($E \sim 10^{44}$~erg) of radiation pulses during the UPE phase. The detailed analysis of the UPE phase of GRB 180720B is presented in \citet{Rastegarnia2022}. Figure~\ref{fig:lightcurve+spectra} shows the light-curve and spectra of different astrophysical processes underlying different episodes in GRB 180720B; for more details see \pone.
  
In this article, we exclusively analyze the radio, optical, and X-ray afterglows of GRB 180720B. We extend the description of the GRB afterglow developed by \citet{2018ApJ...869..101R} within the BdHN scenario. In this approach, the afterglow originates from the synchrotron radiation produced by the expansion of the SN ejecta in the presence of the $\nu$NS magnetic field. We here provide the theoretical formulation of the underlying synchrotron radiation and demonstrate that the spinning $\nu$NS, in addition to the X-ray afterglow, also originates the optical and the radio afterglows. 

In addition, we present the fundamental result that in GRB 180720B, Episode VI, namely, the luminosity of the optical SN bump is expected to be comparable to the optical synchrotron radiation. {Unfortunately, there was no optical follow-up of GRB 180720B at the times} $\sim 15$--$20$ days after the trigger predicted in \citet{2018GCN.23019....1R}; see Fig.~\ref{fig:SN-optical}.

The article is organized as follows. {Section \ref{sec:2} summarizes the sequence of physical phenomena that occur in a BdHN and their related observables in GRB data.} In section \ref{sec:3}, we represent the X-ray, optical and radio afterglow data of GRB 180720B. In section \ref{sec:4}, we represent the formulation related to synchrotron and pulsar radiation of the $\nu$NS. In section \ref{sec:5}, we represent the concluding remarks.

%%%%%%%%%%%%%%%%%%%%%%%%%%%%%%%%%%%%%%%%%%%%%%%%%%%%
%%%%%%%%%%%%%%%%%%%%%%%%%%%%%%%%%%%%%%%%%%%%%%%%%%%%
{
\section{Physical processes and observables of BdHNe}\label{sec:2}
}
%%%%%%%%%%%%%%%%%%%%%%%%%%%%%%%%%%%%%%%%%%%%%%%%%%%%
%%%%%%%%%%%%%%%%%%%%%%%%%%%%%%%%%%%%%%%%%%%%%%%%%%%%

{
We summarize the sequence of physical phenomena in a BdHN, their related electromagnetic emission, and associated observed emission episodes in GRBs data. The BdHN event roots in the induced gravitational collapse (IGC) scenario proposed in \citet{2012ApJ...758L...7R} . It starts with the core-collapse of the CO star that leads to the $\nu$NS at its center, and the ejection of the outermost layers in form of SN explosion. The latter produces a hypercritical accretion process onto the NS companion \cite{2014ApJ...793L..36F, 2016ApJ...833..107B}, while fallback of the innermost layers accrete onto the $\nu$NS \cite{2019ApJ...871...14B}.
}

{
\textbf{\textit{Precursors}.}
The hypercritical accretion onto both NSs can be observed as precursors to the prompt gamma-ray emission of BdHN I (see, e.g., GRB 090618 in \citealp{2016ApJ...833..107B}, GRB 130427A in \citealp{2019ApJ...874...39W}; and also Becerra et al., submitted, for new numerical simulations), or in the prompt emission itself of BdHN II (see, e.g., GRB 180728A in \citealp{2019ApJ...874...39W} and, very recently, GRB 190829A in \citealp{2022arXiv220705619W}).
}

{
From now on, we specialized the discussion on BdHN I. The newborn Kerr BH is surrounded by the magnetic field inherited from the collapsed NS \citep[see][for a discussion on the nature of the magnetic field in BdHN I]{2020ApJ...893..148R} and by low-density ionized matter \citep{2016ApJ...833..107B, 2019ApJ...871...14B, 2019ApJ...883..191R}. These three ingredients conform the so-called {inner engine} of the high-energy emission of the GRB \citep{2019ApJ...886...82R, 2020EPJC...80..300R, 2021A&A...649A..75M, 2021MNRAS.504.5301R}. The BH extractable energy powers both the ultrarelativistic prompt emission (UPE) phase and the GeV emission (see below).
}

{
The gravitomagnetic interaction of the Kerr BH of mass $M$ and angular momentum $J$, with the external magnetic field of strength $B$, induce an electric field in the vicinity of the BH horizon, $E \sim \Omega_H r_H B_0/c = \alpha B_0/2$, where $\Omega_H = c\,\alpha/(2 r_H)$ and $r_H$ are, respectively, the Kerr BH angular velocity and outer horizon. Here, $\alpha = c J/(G M^2)$ is the BH dimensionless spin parameter. The full mathematical expressions of the electromagnetic field outside the Kerr BH for a parallel, asymptotically uniform exterior magnetic field are given by the Papapetrou-Wald solution \citep{1974PhRvD..10.1680W}. The above electric field is overcritical, i.e., $E > E_c = m_e^2 c^3/(e \hbar)$, for magnetic field strengths $B \gtrsim 2 E_c/\alpha$. These values are attainable if the magnetic field is amplified in the process of gravitational collapse of the NS to a BH \citep[see, e.g., discussion in][]{2020ApJ...893..148R}.
}

{
\textbf{\textit{The UPE phase}.} One of the most exciting recent developments is the explanation of the MeV radiation of the UPE phase of GRBs. The necessity of introducing new physics in the GRB prompt emission has arisen from the revealed \textit{hierarchical} or \textit{self-similar} structure observed in the UPE of GRB 190114C \citep{2021PhRvD.104f3043M} and GRB 180720B \citep{Rastegarnia2022}). The quality of the data of the Fermi satellite and Neil Gehrels Swift Observatory, has allowed to perform in-depth a time-resolved analysis of the GRB prompt emission. Such analysis has shown that the spectra of the UPE, on ever decreasing time intervals (of up to a fraction of a second), shows similar BB+CPL spectra. These self-similar BB+CPL spectra on rebinned time intervals point to a microscopic phenomenon operating on shorter and shorter timescales. We are not aware of any explanation of the above UPE hierarchical structure from traditional models based on an ultrarelativistic jet. We have recently explained this phenomenon in the context of the inner engine of BdHN I, for GRB 190114C in \citet{2021PhRvD.104f3043M} and for GRB 180720B in \citet{Rastegarnia2022}. The UPE phase is explained by the overcritical regime of the electric field that leads to the quantum electrodynamics (QED) process of vacuum polarization, i.e., the formation of an electron-positron ($e^+e^-$) pair plasma. The $e^+e^-$ plasma self-accelerates, loads baryons from the medium, and reaches transparency at large radii $R_{\rm tr}$ with Lorentz factor $\Gamma$. \citet{1999A&A...350..334R} presented the first numerical simulations of the dynamics of the expansion and transparency of such optically thick pair electromagnetic-baryon plasma, which has been there called the \textit{PEMB} pulse. The dynamics of the plasma depends only on the initial conditions of energy and baryon load, which following the inner engine theory, depend only on the mass, angular momentum, and electric energy stored in the \textit{dyadoregion}, i.e., the region outside the BH where the electric field is overcritical. Therefore, the plasma dynamics is set by the inner engine parameters $M$, $J$ and $B_0$. Since the electric energy is induced by the gravitomagnetic interaction of the BH and the magnetic field, it is the BH extractable energy that ultimately powers the UPE. Therefore, in each process of expansion and transparency of a PEMB pulse, the Kerr BH loses a fraction of $M$ and $J$. After the energy release of a PEMB pulse, the mass and angular momentum of the BH have decreased to $M = M_0- \Delta M$ and $J = J_0 - \Delta J$. For the UPE of GRB 190114C, we have estimated that the initial PEMB pulse has $R_{\rm tr} \sim 10^9$ cm, $\Gamma \sim 10^2$, and plasma energy $\sim 10^{43}$ erg, leading to $\Delta M/M \sim \Delta J/J\sim 10^{-9}$ \citep{2021PhRvD.104f3043M}. Related to the new lower value of $J$, it corresponds a new lower value of the electric field, $E = E_0 (1-\Delta J/J)$. Therefore, the system restarts a new process characterized by new lower values of $M$, $J$, and $E$, which lead to a lower $e^+e^-$ plasma energy. The extremely short QED timescale of the vacuum polarization process, $\sim \hbar/(m_e c^2) \approx 10^{-21}$ s, guarantees that the process can repeat over time until the electric field reaches the critical value. For GRB 180720B, the UPE phase driven by the above QED mechanism, ends at $t_{\rm rf}=10.89$~s.
}

{
\textbf{\textit{The Cavity}.} The SPH simulations of the SN explosion and the NS accretion show the region around the NS that collapses to a BH is characterized by low density of the order of $10^{-6}$ g cm$^{-3}$. Numerical simulations show that the density inside this cavity is further depleted by the BH formation and the expansion of the $e^+e^-$ plasma to values of the order of $\sim 10^{-14}$--$10^{-13}$ g cm$^{-3}$ \citep{2019ApJ...883..191R}. The emission from the cavity walls, characterized by a CPL spectrum, has been there identified in the case of GRB 190114C. The cavity emission has been also identified in GRB 180720B (see Fig. \ref{fig:lightcurve+spectra} and Table \ref{tab:Summary}).
}

{
\textbf{\textit{The soft and hard X-ray flares (SXFs and HXFs)}.} They originate from the \textit{breakout} of the $e^+e^-$ plasma from high-density regions of the ejecta around the BH. In those regions, the $e^+e^-$ plasma loads more baryons than in the case of the low-density regions leading to the UPE phase. This leads the plasma to reach transparency with a lower Lorentz factor $\lesssim 5$ \citep{2018ApJ...852...53R}. The occurrence of this emission in several GRBs has been analyzed in light of the numerical simulations of the $e^+e^-$ plasma expansion in the realistic SN ejecta obtained from the three-dimensional simulations, and it has been shown that the emission is viewed within at an intermediate angle, between the binary plane and the rotation axis \citep[see, also,][]{2021MNRAS.504.5301R}. 
}

{
\textbf{\textit{The GeV emission}.}
The induced electric field accelerates to ultrarelativistic energies electrons surrounding the BH. Along the BH rotation axis, the electric and magnetic field are parallel and the electrons accelerate without energy losses. Under these conditions, the electrons reach energies of up to $10^{18}$~eV and protons of up to $10^{21}$~eV \cite{2020EPJC...80..300R}, leading to ultra high-energy cosmic rays (UHECRs). Outside the polar axis, the synchrotron radiation of the accelerated electrons peaks at GeV energies, explaining the observed GeV radiation of GRBs \cite{2019ApJ...886...82R, 2020EPJC...80..300R, 2021A&A...649A..75M, 2022ApJ...929...56R}. The theoretical framework of the inner engine based on the above description has been applied to the explanation of the GeV emission of GRBs. The case of GRB 130427A has been analyzed in \citet{2019ApJ...886...82R}, GRB 190114C and the extension of the theoretical framework to active galactic nuclei (AGNs) like M87* in \citet{2020EPJC...80..300R, 2021A&A...649A..75M}. A rigorous general relativistic treatment of the inner engine and the emission produced by charged particles accelerated by the induced electric field can be found in \citet{2022ApJ...929...56R}. The theoretical model predicts that the above emission process occurs within $60^\circ$ degrees from the BH rotation axis, which agrees with the lack of observed GeV emission in a fraction of BdHN I \citep[see][for details]{2021MNRAS.504.5301R}.
}

{
\textbf{\textit{The multiwavelength afterglow}.} This is the main topic of this article. In the BdHN model, the afterglow originates from synchrotron radiation in the SN ejecta, powered by the $\nu$NS and not from ultrarelativistic blastwaves \cite{2018ApJ...869..101R, 2019ApJ...874...39W, 2020ApJ...893..148R}. These works have carried out the numerical integration of the electron kinetic equation taking into account the expansion of the SN from numerical simulations, the radial dependence of the magnetic field expected from pulsar theory, and the power of the $\nu$NS as a pulsar. More recently, we have presented analytic solutions of the above treatment that are consistent with our previous numerical calculations. \citet{2022arXiv220705619W} has presented the analytic treatment of the X-ray, radio and optical afterglow emission of GRB 190829A. The consistency of this theoretical treatment with the multiwavelength afterglow data is here extended to the case of GRB 180720B.
}

{
\textbf{\textit{The optical SN}.}
We fnish the general description of BdHNe with the emission of the SN in the optical wavelengths. This emission is powered by the energy release of nickel decay (into cobalt) in the ejecta. The SN associated with GRBs are similar with each other, roughly independent of the wide GRB energetics (see, e.g., \citealp{2017AdAst2017E...5C} and Aimuratov et al., to be submitted). This indicates that the pre-SN progenitor (i.e., the CO star) leading to the core-collapse SN event triggering the GRB, is similar in all GRBs. This is a relevant information for the formation channel of the CO-NS binaries leading to GRBs (see, e.g., \citealp{2015PhRvL.115w1102F}).
}

%%%%%%%%%%%%%%%%%%%%%%%%%%%%%%%%%%%%%%%%%%%%%%%%%%%%
%%%%%%%%%%%%%%%%%%%%%%%%%%%%%%%%%%%%%%%%%%%%%%%%%%%%
\section{Data of GRB 180720B}\label{sec:3}
%%%%%%%%%%%%%%%%%%%%%%%%%%%%%%%%%%%%%%%%%%%%%%%%%%%%
%%%%%%%%%%%%%%%%%%%%%%%%%%%%%%%%%%%%%%%%%%%%%%%%%%%%

On 20 July 2018, GRB 180720B triggered Fermi-GBM at 14:21:39.65 UT \citep{2018GCN.22981....1R}, Swift-BAT at 14:21:44 UT \citep{2018GCN.22973....1S}, and Fermi-LAT at 14:21:44.55 UT \citep{2018GCN.22980....1B}. Swift-XRT began to observe $91$ seconds after the Fermi-GBM trigger \citep{2018GCN.22973....1S}. These gamma-ray and X-ray data are retrieved and analyzed by Fermi GBM Data Tools \citep{GbmDataTools} and HEASoft\footnote{\url{https://heasarc.gsfc.nasa.gov/lheasoft/}} respectively. The corresponding gamma-ray and X-ray luminosity light-curves are shown in Fig. \ref{fig:lightcurve}. This GRB also triggered several optical telescopes. The optical light-curve of r-band in Fig. \ref{fig:lightcurve} is reproduced from \citet{2019Natur.575..464A}, which gathers the observations from Kanata~\citep{2018GCN.22977....1S}, MITSuME~\citep{2018GCN.22983....1I}, TSHAO~\citep{2018GCN.22979....1R}, MASTER~\citep{2018GCN.23023....1L}, ISON-Castelgrande~\citep{2018GCN.23020....1S}, OSN~\citep{2018GCN.22985....1K}, LCO~\citep{2018GCN.22976....1M}, and KAIT~\citep{2018GCN.23033....1Z}.

\begin{figure}
\centering
\includegraphics[width=\hsize,clip]{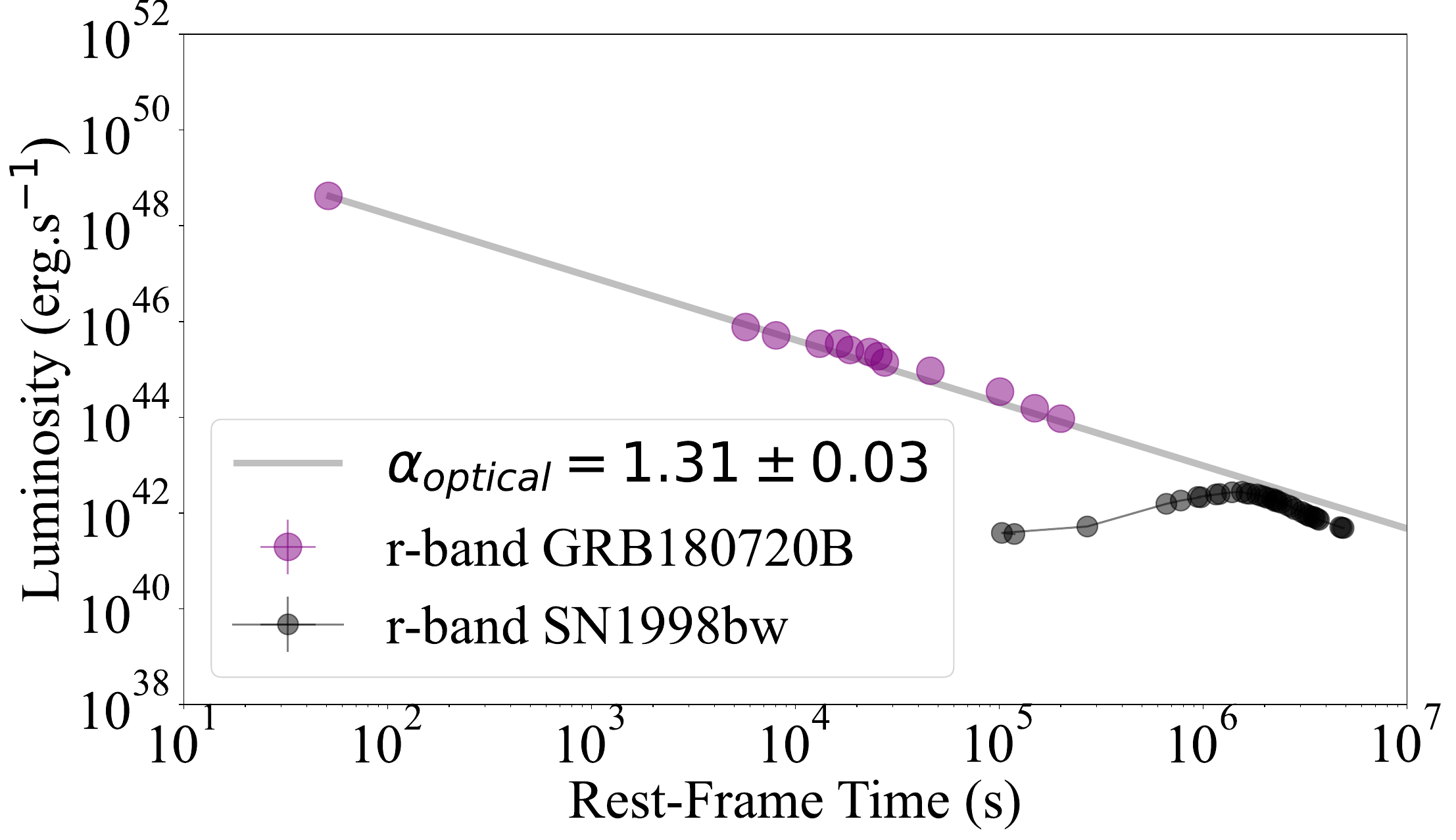}
\caption{The optical luminosity of prototype SN 1998bw (grey) vs. the optical (purple) afterglow of GRB 180720B. In this GRB, no optical data were observed at the time of $\sim 16$--$24$ day after the trigger as we predicted in \citet{2018GCN.23019....1R}. Most likely, the associated SN could have been detected if optical observations had been made at the time. The r-band optical data retrieved from \citet{2019Natur.575..464A}. The optical luminosity of prototype SN 1998bw is plotted from reference \citep{1998Natur.395..672I}.}
\label{fig:SN-optical}
\end{figure}

\begin{figure*}
\centering
\includegraphics[width=0.85\hsize,clip]{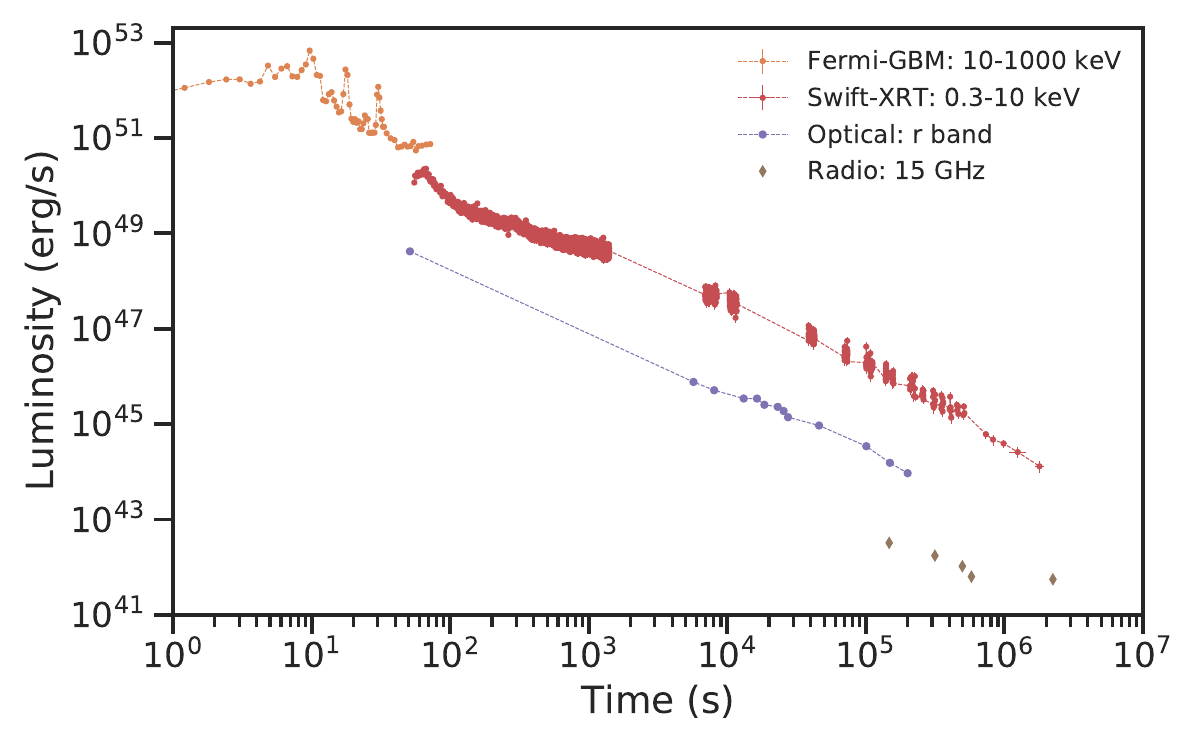}
\caption{Luminosity light-curve of GRB 180720B: including the observations of gamma-ray from Fermi-GBM (yellow), soft X-ray from Swift-XRT (red), the brown diamonds present the radio data from AMI-LA retrieved from \citet{2020MNRAS.496.3326R}, the purple circles represent the r-band optical data retrieved from \citet{2019Natur.575..464A}.}
\label{fig:lightcurve}
\end{figure*}

KAIT provides the latest observation at $3.874$ days after the trigger. This time is much {before than} the time {expected} of SN optical bump, based on the previous SNe statistics \citep{2017AdAst2017E...5C}. Considering the burst occurs at a distance of $z = 0.654$, \citet{2018GCN.23019....1R} predicted that the SN optical flux {should have peaked} at $21.8 \pm 4.3$ days after the trigger{, i.e., on August 11, 2018, with the uncertainty from August 7 to 15, 2018}. In figure \ref{fig:SN-optical}, {we plot the optical luminosity of} SN 1998bw {and the observed optical afterglow of GRB 180720B.} The peak luminosities of other {SNe associated with GRBs} vary about $0.5$ to $3$ times \citep{2017AdAst2017E...5C}. The {figure shows that the} expected optical flux of the SN is not much lower than the optical flux of the synchrotron radiation. {Therefore,} the SN signal could {have been} observed if the {source were monitored at the optical wavelengths at $21.8 \pm 4.3$ days, as suggested in \citet{2018GCN.23019....1R}. Regretfully, we have to conclude that} the missing of SN optical signal is due to the missed prolonged optical {follow-up of GRB 180720B (see Fig.~\ref{fig:SN-optical}), which has been a big missing opportunity for a further observational verification of a theoretical prediction of the BdHN model.}

The GRB 180720B triggered also radio emission observed by Arcminute Microkelvin Imager Large Array (AMI-LA) \citep{10.1093/mnras/sts259,10.1093/mnras/stx2407}. The observed flux which have been made over $5$ logarithmically spaced data points, were retrieved from \citep{2020MNRAS.496.3326R}.

The broad-band luminosity light-curve of GRB 180720B is represented in Fig.~\ref{fig:lightcurve}. 

\begin{table*}
\centering
\caption{The episodes and afterglows of GRB 180720B. This table reports the name, the underlying astrophysical process, the duration (s), the best fit spectrum, the isotropic energy (erg), and the reference for each event in GRB 180720B.}
\label{tab:Summary}
\small\addtolength{\tabcolsep}{-3pt}
\begin{tabular}{|c|c|c|c|c|c|c}

\hline
Event& Astrophysical Process & duration(s) & Spectrum & $E_{\rm iso}$ (erg) & Ref \\
\hline
\hline

&&&CPL+BB&&\\
I) BH--rise& Vacuum polarization&1.21&$\alpha=-1.13$, $E_{\rm c}=2220.57$~keV&$(6.37\pm 0.48) \times 10^{52}$&\citet{2021PhRvD.104f3043M}\\
(UPE I)&around BH&&$kT=50.31$~keV&&\\
\hline
&&&CPL&&\\
II) $\nu$NS--rise& Synchrotron radiation&3.01&$\alpha=-0.98$, $E_{\rm c}=737$~keV &$(1.13\pm0.04) \times 10^{53}$&\citet{2020ApJ...893..148R}\\
& powered by $\nu$NS&&&&\\
\hline
&&&CPL&&\\
III) BH--rise& Synchrotron radiation&600&$\alpha = -2.0 \pm 0.1$&$(2.2\pm 0.2) \times 10^{52}$&\citet{2019ApJ...886...82R}\\
(GeV radiation)&powered by BH&&&&\\

&&&CPL+BB&&\\
 BH--rise& Vacuum polarization&1.82&$\alpha= -1.06$, $E_{\rm c}=1502.5$~keV&$(1.6\pm 0.95) \times 10^{53}$&\citet{2021PhRvD.104f3043M}\\
(UPE II)&around BH&&$kT= 39.8$~keV&&\\
\hline
&&&CPL&&\\
IV) Cavity& Radiation from cavity's&3.02&$\alpha= -1.16$, $E_{\rm c}=607.96$~keV&$(4.32\pm 0.19) \times 10^{52}$&\citet{2019ApJ...883..191R}\\
&wall&&&&\\
\hline
&&&CPL&&\\
V) HXF& Emission from outer&6.03&$\alpha=-1.198 \pm 0.031$&$(3.93\pm 0.33) \times 10^{52}$&\citet{2018ApJ...852...53R}\\
&SN layers&&$E_{\rm c} = 550 $~keV&&\\
\hline
&&&CPL+BB&&\\
VI) SXF& Emission from outer&15.12&$\alpha=-1.79 \pm 0.23$&$(2.89\pm 0.42) \times 10^{52}$&\citet{2018ApJ...869..151R}\\
&SN layers&&$kT=0.99 \pm 0.13$~keV&&\\
\hline 
&&expected to be&&expected to be &\\
VII) Optical SN& Nickel decay &$\sim 10^6 $&PL&$ \sim 10^{49}$&Not observed\\
& &&&& for GRB 180720B \\
\hline
&&&PL&&\\
X-ray Afterglow& Synchrotron radiation&$\sim 10^7$&&$(2.61\pm 1.0) \times 10^{52}$&\citet{2020ApJ...893..148R}\\
& powered by $\nu$NS &&&&\\
 \hline 
&&&PL&&\\
Optical Afterglow& Synchrotron radiation&$\sim 3\times 10^5$&&$(6.10\pm 1.0) \times 10^{50}$&\citet{2019Natur.575..464A}\\
& powered by $\nu$NS &&&&This work\\
 \hline 
 &&&PL&&\\
sub--TeV emission& Unknown&$\sim 3\times 10^3$&&$(2.40\pm 1.8) \times 10^{50}$&\citet{2019Natur.575..464A}\\
&&&&&\\
\hline 
&&&PL&&\\
Radio Afterglow& Synchrotron radiation&$\sim 2 \times 10^6$&&$(2.21\pm 0.24) \times 10^{46}$&\citet{2020MNRAS.496.3326R}\\
& powered by $\nu$NS &&&&This work\\
\hline 

\end{tabular}
\end{table*}

%\textcolor{blue}{(The plot file and data are here, in case liang or someone needs to update the figure: \url{http://54.38.124.159:5888/notebooks/180720B/180720B.ipynb})} 

%%%%%%%%%%%%%%%%%%%%%%%%%%%%%%%%%%%%%%%%%%%%%%%%%%%%
%%%%%%%%%%%%%%%%%%%%%%%%%%%%%%%%%%%%%%%%%%%%%%%%%%%%
\section{Synchrotron and pulsar radiation of the $\nu$NS}\label{sec:4}
%%%%%%%%%%%%%%%%%%%%%%%%%%%%%%%%%%%%%%%%%%%%%%%%%%%%
%%%%%%%%%%%%%%%%%%%%%%%%%%%%%%%%%%%%%%%%%%%%%%%%%%%%

We here follow and extend the treatment of the GRB afterglow by \citet{2018ApJ...869..101R} within the BdHN scenario. In this picture, the afterglow originates from the synchrotron radiation produced by the expansion of the SN ejecta in presence of the magnetic field of the $\nu$NS. We now estimate the emission generated by the synchrotron mechanism in the X-rays, in the optical, and in the radio, together with the pulsar emission of the $\nu$NS.

%%%%%%%%%%%%%%%%%%%%%%%%%%%%%%%%%%%%%%%%%%%%%%%%%%%%%
\subsection{Synchrotron emission by the expanding ejecta}\label{sec:synch1}
%%%%%%%%%%%%%%%%%%%%%%%%%%%%%%%%%%%%%%%%%%%%%%%%%%%%%

The evolution of the distribution of radiating electrons per unit energy, $N(E,t)$, is determined by the kinetic equation \citep[see, e.g.,][]{1962SvA.....6..317K, 1979rpa..book.....R}
\begin{equation}\label{eq:kinetic}
    \frac{\partial N(E, t)}{\partial t}=-\frac{\partial}{\partial E}\left[\dot{E}\,N(E,t)\right] + Q(E,t),
\end{equation}
where $Q(E,t)$ is the number of injected electrons into the ejecta per unit time $t$, per unit energy $E$, and $\dot E$ is the electron energy loss rate. 

In the present model, the electrons are subjected to adiabatic losses due to the ejecta expansion and synchrotron radiation losses because of the presence of the $\nu$NS magnetic field. The electrons lose their energy efficiently by synchrotron radiation, so we can assume a one-zone model adopting that the emission originates from the innermost layer of the ejecta, which we denote as $r=R_*$. We assume that the ejecta expand at constant velocity $v_{*,0}$, so the radius evolves as
\begin{equation}\label{eq:radius}
    R_*(t) = R_{*,0}\,\hat{t},
\end{equation}
where $\hat{t} \equiv t/t_*$, and $t_* \equiv R_{*,0}/v_{*,0}$.

Having specified the pace at which the ejecta expand, we can write the energy balance equation governing the evolution of the electron's energy \citep[see, e.g.,][]{1962SvA.....6..317K}
\begin{equation}\label{eq:gammadot}
    -\dot E = \frac{E}{t} + P_{\rm syn}(E,t),
\end{equation}
where the first term on the right-hand side accounts for expansion, adiabatic losses, and the second term is given by the bolometric synchrotron radiation power \citep[see, e.g.,][]{2011hea..book.....L}
\begin{equation}\label{eq:Psyn}
    P_{\rm syn}(E,t) = \beta B_*^2(t) E^2,
\end{equation}
where $\beta = 2e^4/(3 m_e^4 c^7)$, and $B_*(t)$ is the magnetic field the electrons are subjected to. From the traditional pulsar theory, we expect that beyond the light-cylinder, $r=c/\Omega$, where $\Omega$ is the rotation angular velocity of the $\nu$NS, the magnetic field of the $\nu$NS decreases linearly with distance \citep[see, e.g.,][]{1969ApJ...157..869G,1969ApJ...157.1395O}. Therefore, we adopt that the magnetic field strength at the ejecta position varies with time as
\begin{equation}\label{eq:B}
    B_*(t) = B_{*,0}\, \frac{R_{*,0}}{R_*} = \frac{B_{*,0}}{\hat{t}},
\end{equation}
where $B^{(0)}_i$ is the magnetic field strength at $r=R_{*,0}$, and we have used Eq. (\ref{eq:radius}).

For completing the specification of Eq. (\ref{eq:kinetic}), we need to introduce the distribution of the injected electrons. We assume the power-law distribution \citep[see, e.g.,][]{1962SvA.....6..317K, 1973ApJ...186..249P, 1979rpa..book.....R, 2011hea..book.....L}
\begin{equation}\label{eq:Q}
Q(E,t)=Q_0(t)E^{-\gamma}, \qquad 0\leq E \leq E_{\rm max},
\end{equation}
where $\gamma$ and $E_{\rm max}$ are parameters to be determined from the observational data, and $Q_0(t)$ can be related to the power released by the $\nu$NS and injected into the ejecta. We adopt an injected power of the form
\begin{equation}\label{eq:Lt}
L_{\rm inj}(t) = 
L_0 \left(1+\frac{t}{t_q}\right)^{-k},
\end{equation}
where $L_0$, $t_q$, and $k$ are model parameters to be obtained from the fit of the observational data. Therefore, the function $Q_0(t)$ can be found from
\begin{align}\label{eq:LandQ}
L_{\rm inj}(t) &= \int_{0}^{E_{\rm max}} E\,Q(E,t) dE =Q_0(t)\frac{E_\mathrm{max}^{2-\gamma}}{2-\gamma},
\end{align}
which using Eq.~(\ref{eq:Lt}) leads to
\begin{equation}\label{eq:Q0}
    Q_0(t) =
q_0\left(1+\frac{t}{t_q}\right)^{-k},
\end{equation}
where $q_0 \equiv  (2-\gamma)L_0/E_{\rm max}^{2-\gamma}$.

Having specified the evolution of the ejecta by Eq.~(\ref{eq:radius}) and the magnetic field by Eq.~(\ref{eq:B}), as well as the rate of particle injection given by Eqs.~(\ref{eq:Q}) and (\ref{eq:Q0}), we can now proceed to the integration of the kinetic equation (\ref{eq:kinetic}).

First, we find the evolution of a generic electron injected at time $t=t_i$ with energy $E_i$. Equation~(\ref{eq:gammadot}) can be integrated analytically leading to the energy evolution \citep{2022arXiv220200316R, 2022arXiv220200314R}
\begin{equation}\label{eq:gammavst}
    E = \frac{E_i\,(t_i/t)}{1 + {\cal M} E_i t_i\left( \frac{1}{\hat{t}_i^{2}} -  \frac{1}{\hat{t}^{2}}\right)},
\end{equation}
where ${\cal M}\equiv \beta B^2_{*,0}/2$.

The solution of Eq. (\ref{eq:kinetic}) can be written as \citep[see, e.g.,][]{1973ApJ...186..249P}
\begin{equation}\label{eq:Nsol}
    N(E,t) = \int_E^\infty Q[E_i, t_i(t,E_i,E)] \frac{\partial t_i}{\partial E} dE_i,
\end{equation}
where $t_i(t,E_i,E)$ is obtained from Eq. (\ref{eq:gammavst}).

We can write $N(E,t)$ as a piecewise function of time depending upon the behavior of the energy injection (\ref{eq:Q0}). All the observational data of GRB 190114C is contained in the regime where synchrotron losses are dominant. In this case, the solution of Eq. (\ref{eq:Nsol}) is well approximated by \citep{2022arXiv220200316R}
\begin{align}\label{eq:N3}
&N(E,t)\approx \begin{cases}
    \frac{q_0}{\beta B_{*,0}^2 (\gamma-1)}\hat{t}^{2} E^{-(\gamma+1)}, & t < t_q\\
   \frac{q_0 (t_q/t_*)^{k}}{\beta B_{*,0}^2 (\gamma-1)}\hat{t}^{2-k} E^{-(\gamma+1)}, &   t_q < t < t_b,
\end{cases}
\end{align}
where $E_b < E < E_{\rm max}$, being
\begin{equation}\label{eq:Eb}
    E_b = \frac{\hat{t}}{{\cal M} t_*},\quad
    t_b = t_*^2 {\cal M}  E_{\rm max}.
\end{equation}

The electrons emit most of the synchrotron radiation at frequencies close to the critical frequency $\nu_{\rm crit} = \alpha B_* E^2$, where $\alpha = 3 e/(4\pi m_e^3 c^5)$. Therefore, we can assume $\nu \approx \nu_{\rm crit}$, so the bolometric synchrotron power (\ref{eq:Psyn}) can be written approximately in terms of the radiation frequency as
\begin{equation}\label{eq:Psynbol}
    P_{\rm syn}(E,t) \approx P_{\rm syn}(\nu,t) = \frac{\beta}{\alpha} B_* \nu,
\end{equation}
and the spectral density, i.e. energy per unit time, per unit frequency, as \citep[see, e.g.,][]{2011hea..book.....L}
\begin{align}\label{eq:Jnu1}
    J_{\rm syn}(\nu,t) &\approx P_{\rm syn}(\nu,t) N(E,t) \frac{dE}{d\nu} \nonumber \\
    &= \frac{\beta\ \eta}{2} \alpha^{\frac{p-3}{2}} B_{*,0}^{\frac{p+1}{2}}\hat{t}^{\frac{2 l-p-1}{2}}\nu^{\frac{1-p}{2}}.
\end{align}
where we have used $N(E,t) = \eta\,\hat{t}^l E^{-p}$, being $\eta$ , $l$ and $p$, known constants from Eq. (\ref{eq:N3}).

The synchrotron luminosity radiated in the frequencies $[\nu_1,\nu_2]$ can be then obtained as
\begin{equation}\label{eq:Lnu}
    L_{\rm syn}(\nu_1,\nu_2; t) = \int_{\nu_1}^{\nu_2} J_{\rm syn}(\nu,t)d\nu\approx \nu J_{\rm syn}(\nu,t),
\end{equation}
where $\nu_1=\nu$, $\nu_2=\nu+\Delta\nu$, being $\Delta\nu$ the bandwidth. Because $J_{\rm syn}$ shows a power-law behavior in frequency (see Eq. \ref{eq:Jnu1}), we have used in the second equality of Eq. (\ref{eq:Lnu}) the approximation $\Delta\nu/\nu\ll 1$. By substituting Eq. (\ref{eq:Jnu1}) into Eq. (\ref{eq:Lnu2}), we obtain that, at the frequency $\nu$, the synchrotron luminosity is given by
\begin{equation}\label{eq:Lnu2}
    L_{\rm syn}(\nu, t) = \frac{\beta}{2} \alpha^{\frac{p-3}{2}} \eta B_{*,0}^{\frac{p+1}{2}}\hat{t}^{\frac{2 l-p-1}{2}}\nu^{\frac{3-p}{2}}.
\end{equation}

Equation (\ref{eq:Lnu2}) implies that the synchrotron radiation follows a power-law behavior both in time and radiation frequency, $L_{\rm syn} \propto t^{\frac{2 l - p-1)}{2}} \nu^{\frac{3-p}{2}}$, where $p = \gamma+1$ and the value of $l$ depend on whether we are in the phase of constant injection or power-law injection (see Eq. \ref{eq:N3}). Therefore, the synchrotron radiation leads to a \textit{rainbow} luminosity, characterized by the same power-law index in the X-rays, optical, and radio wavelengths (see Fig. \ref{fig:fit180720B}). This occurs when the system remains over time in the same physical regime, namely when the observational data is contained within times $t<t_b$, and the electron energies are in the range $E_b<E<E_{\rm max}$, see Eq. (\ref{eq:Eb}). Otherwise, the power-law behavior changes when the system makes a transition to a regime of dominance of adiabatic over synchrotron losses. In the present case, we did not find evidence in the GRB afterglow data of the occurrence of the transition to such a physical regime. This implies that the ratio of the luminosity at different frequencies depends only on the power-law index of the injection rate as \citep{2022arXiv220200316R}
\begin{equation}\label{eq:Lratio}
    \frac{L_{\rm syn} (\nu_1)}{L_{\rm syn} (\nu_2)} = \left( \frac{\nu_1}{\nu_2} \right)^{\frac{3-p}{2}} = \left( \frac{\nu_1}{\nu_2} \right)^{\frac{2-\gamma}{2}}.
\end{equation}
Therefore, we can set the value of $\gamma$ by fitting the observed X-rays to optical luminosity ratio. Once we have fixed $\gamma$, the optical (or X-rays) to radio luminosity ratio is also set. Figure \ref{fig:fit180720B} shows that the obtained synchrotron luminosity at radio wavelengths also agrees with the observational data. This result implies that the model leads to the correct observed spectrum of the afterglow emission in the wide range of energies from the radio to the X-rays, which strongly support the present scenario.

%%%%%%%%%%%%%%%%%%%%%%%%%%%%%%%%%%%%%%%%%%%%%%%%%%%
%%%%%%%%%%%%%%%%%%%%%%%%%%%%%%%%%%%%%%%%%%%%%%%%%%%
\subsection{Newborn NS evolution and pulsar emission}\label{sec:synch2}
%%%%%%%%%%%%%%%%%%%%%%%%%%%%%%%%%%%%%%%%%%%%%%%%%%%
%%%%%%%%%%%%%%%%%%%%%%%%%%%%%%%%%%%%%%%%%%%%%%%%%%%

At some point in the time evolution, when the synchrotron luminosity has fallen sufficiently, the pulsar emission of the $\nu$NS becomes observable. We calculate this emission by adopting a dipole+quadrupole magnetic field model following \citet{2015MNRAS.450..714P}. In this model, the total pulsar (spindown) luminosity is given by
\begin{align}\label{eq:Lsd}
    L_{\rm sd} &= L_{\rm dip} + L_{\rm quad} \nonumber \\
    &= \frac{2}{3 c^3} \Omega^4 B_{\rm dip}^2 R^6 \left( 1 + \xi^2 \frac{16}{45} \frac{R^2 \Omega^2}{c^2} \right),
\end{align}
with $R$ the $\nu$NS radius, and $\xi$ defines the quadrupole to dipole strength ratio
\begin{equation}
    \xi \equiv \sqrt{\cos^2\chi_2+10\sin^2\chi_2} \frac{B_{\rm quad}}{B_{\rm dip}},
\label{eq:eta}
\end{equation} 
{where the angles $\chi_1$ and $\chi_2$ define the quadrupole field geometry, and we have assumed the dipole field in the $m=1$ mode, i.e., inclined $90^\circ$ with respect to the stellar rotation axis. The quadrupole field span between the $m=0$ and $m=1$ modes for values of the angles in the range $0^\circ$--$90^\circ$ \citep[see][for further details]{2015MNRAS.450..714P}.} 

The evolution of the $\nu$NS is calculated from by integrating the energy balance equation
\begin{equation}\label{eq:Erot}
	-(\dot{W}+\dot{T}) = L_{\rm tot} = L_{\rm inj} + L_{\rm sd},
\end{equation}
where $W$ and $T$ are, respectively, the $\nu$NS gravitational and rotational energy. 

\begin{table}
    \centering
    \begin{tabular}{l|r}
    Parameter & Value \\
    \hline
      $\gamma$ &  $1.34$\\
       $k$  & $1.73$\\
       $L_0$ ($10^{50}$ erg s$^{-1}$)& $5.50$\\
       $E_{\rm max}$ ($10^4 \ m_e c^2$) & $1.00$\\
       $t_q$ (s) & $40.00$\\
       $R_{*,0}$ ($10^{10}$ cm) & $1.99$ \\
       $v_{*,0}$ ($10^{9}$ cm s$^{-1}$) & $1.00$ \\
       $B_{*,0}$ ($10^{6}$ G) & $2.51$\\
       $\xi$ & $200.00$\\
       $B_{\rm dip}$ ($10^{13}$ G) & $1.50$ \\
       $P$ (ms) & $1.00$\\
       \hline
    \end{tabular}
    \caption{Numerical values of the theoretical model of synchrotron radiation that fit the multiwavelength observational data of GRB 180720B as shown in Fig. \ref{fig:fit180720B}.}
    \label{tab:parameters}
\end{table}

\begin{figure*}
\centering
\includegraphics[width=0.85\hsize,clip]{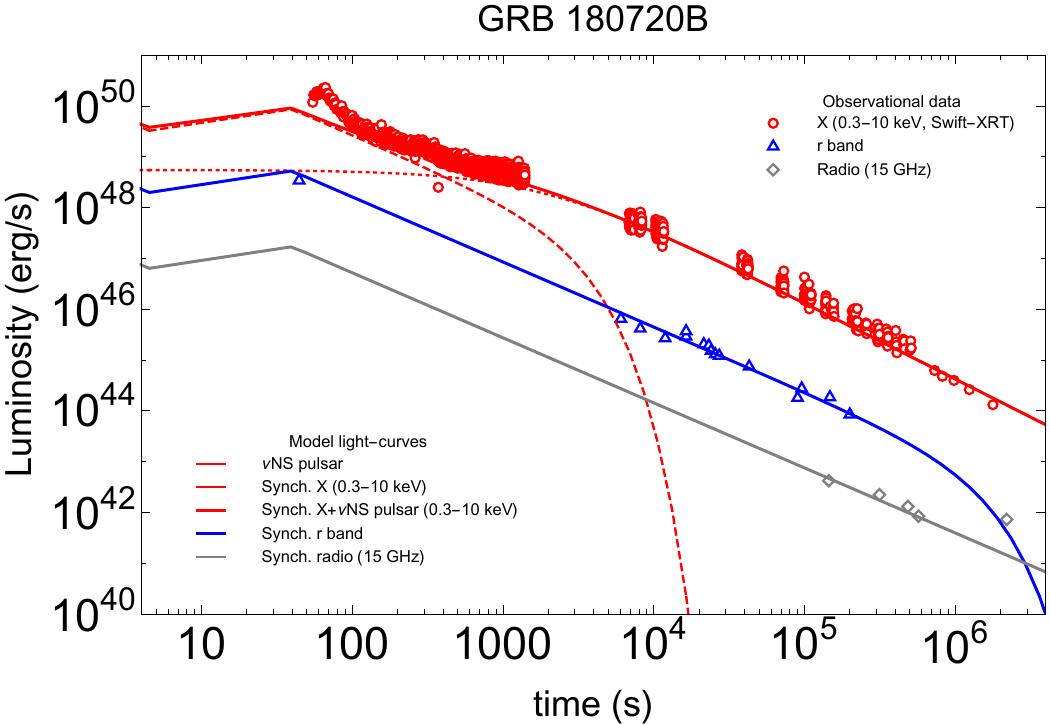}
\caption{The luminosity of GRB 180720B in the X-rays ($0.3$--$10$ keV), optical (r band), and radio ($5.0$ and $9.0$ GHz) energy bands. The X-ray data are produced by the methods developed by the Swift-XRT team to produce the the light curves \citep{2009MNRAS.397.1177E}. The radio data from AMI--LA are retrieved from \citet{2020MNRAS.496.3326R}, the r-band optical data are retrieved from \citet{2019Natur.575..464A}.}
\label{fig:fit180720B}
\end{figure*}

Table \ref{tab:parameters} lists the values of the model parameters that fit the afterglow of GRB 180720B in the X-rays, optical, and radio energy bands. Figure \ref{fig:fit180720B} shows that the observed afterglow luminosity that fades with time with the same power-law behavior in the X-rays, optical and radio is explained by the synchrotron emission. The synchrotron luminosity in the X-rays decays exponentially after a few $10^2$ s, while it keeps its power-law behavior at lower energies. This occurs because around this time the critical synchrotron radiation energy ($h \nu_{\rm crit}$) falls below the keV range. After this time, the pulsar emission from the $\nu$NS dominates the X-ray emission. {The pulsar emission is characterized by a plateau followed by a power-law decay (at times longer than the characteristic spindown timescale). However, in the afterglow there is also the additional power-law contribution from the synchrotron emission. When the plateau phase of the pulsar emission is comparable (but smaller) to the synchrotron power-law luminosity, the sum of the two contributions can lead to a power-law luminosity shallower than the power-law of the pure synchrotron radiation. In GRB 180720B, the X-ray afterglow shows two different power-laws, the first at times $10^2$--$10^3$ s, and the other at times $>10^4$ s (there is a hole of data at $10^3$--$10^4$ s). In the time interval $10^2$--$10^3$ s, the X-ray luminosity shows a shallower power-law than the pure synchrotron luminosity, as it can be seen by comparing it with the power-law of the optical and radio synchrotron at times $>10^4$ s. Such a shallower power-law luminosity indicates the presence of the $\nu$NS magnetic-braking radiation, and indeed it is well fitted by the sum of the synchrotron and the plateau of the newborn NS pulsar emission (see Fig. \ref{fig:fit180720B}).}

The subsequent dominant role of the pulsar emission in the observed X-rays emission has allowed us to infer the strength of the dipole and the quadrupole components of the magnetic field, as well as the rotation period of the $\nu$NS. {Therefore, in the BdHN model the afterglow is characterized by a typical power-law luminosity in the X-rays, optical and radio wavelengths, given by the synchrotron radiation emitted by the expanding SN ejecta in the magnetized medium of the $\nu$NS, together with the pulsar-like emission of the latter.}

The radio emission shows an excess over the synchrotron emission at a few $10^6$ s whose nature is unclear. This excess could be a signature of the SN or the emergence of the $\nu$NS pulsar in the radio wavelengths, but further observational data at longer times and additional theoretical modeling is needed to confirm or disregard any of these possibilities.

%%%%%%%%%%%%%%%%%%%%%%%%%%%%%%%%%%%%%%%%%%%%%%%%%%%%
%%%%%%%%%%%%%%%%%%%%%%%%%%%%%%%%%%%%%%%%%%%%%%%%%%%%
\section{Conclusions}\label{sec:5}
%%%%%%%%%%%%%%%%%%%%%%%%%%%%%%%%%%%%%%%%%%%%%%%%%%%%
%%%%%%%%%%%%%%%%%%%%%%%%%%%%%%%%%%%%%%%%%%%%%%%%%%%%

We have pointed out the essential role of the gravitational collapse of the CO star of about $10 M_\odot$ in presence of a companion NS leading to a variety of independent processes: 1) the SN originated from the CO$_{\rm core}$ collapse, 2) the ejecta accreting on the $\nu$NS spins it up to milliseconds and generates the radio, optical and X-ray afterglows, 3) the ejecta accreting onto the companion NS, leading to the BH formation that generates the UPE phase and GeV emission.

We have here addressed the afterglows driving the three different components of X-ray, radio and optical emissions. {We have shown that the synchrotron emission generated by accelerated electrons in the expanding SN ejecta, powered by the $\nu$NS injected energy, explain the observed power-law emission in the X-rays, optical and radio wavelengths. In our interpretation, the synchrotron power-law luminosity in the X-rays ends about $\sim 10^2$ s due to the transition of the synchrotron critical frequency to values below the X-ray range. At $10^2$--$10^3$ s, the X-ray luminosity shows a shallower decay that resembles a plateau, which we have interpreted as the emergence of the $\nu$NS pulsar emission. The fit of such behavior of the X-ray luminosity and its subsequent power-law emission with the pulsar emission has allowed us to infer the $\nu$NS initial spin ($1$ ms), and the dipole and quadrupole magnetic field strengths, respectively, $B_{\rm dip} = 1.5\times 10^{13}$ G and $B_{\rm quad} = 200 B_{\rm dip}$ (see Table \ref{tab:parameters} for the values of all the model parameters).}

The synchrotron radiation in the optical wavelengths overcomes most of the optical SN except for the SN peak that could have been barely observed. However, no observational data {of GRB 180720B} were acquired at the needed time of $\sim 20$ day after the trigger, {where} we predicted {the occurrence of the SN peak time} \citep{2018GCN.23019....1R}. The associated SN could have been detected if observations at optical wavelengths had been taken at those times.

From the analysis presented in \pone, the most general BdHN I, GRB 180720B, comprises seven independent Episodes characterized  by specific spectral signatures and energetics. They originate from the gravitational collapse of a CO star in the presence of a binary NS companion. Their successful analysis has clarified the occurrence of the most energetic and possibly one of the most complex systems in the Universe. Some Episodes have been observed in selected BdHNe I, e.g., in GRB 190114C \citep{2021MNRAS.504.5301R}, and GRB 130427A \citep{2019ApJ...886...82R}. It is the first time that six of these Episodes have been observed in a single source. This has been made possible by the statistically significant data available and the special inclination of the viewing angle with respect to the equatorial plane of the binary system in GRB 180720B. The latter made possible the observation of the HXF and the SXF. Only the seventh Episode, the final observation of the optical SN due to the nickel decay, with a predicted peak bolometric optical luminosity of $L_{\rm p,avg}= (8.9\pm 3.8 ) \times 10^{42}$ erg s$^{-1}$ and rest-frame peak time of $t_{\rm p,avg}= (1.2\pm 0.24) \times 10^{6}$ s, was observationally missed.

In summary, the interaction of the SN ejecta with the$\nu$NS originated the emission of $(1.39\pm 0.05) \times 10^{53}$~erg, revealed in the $\nu$NS-rise, and different afterglows in the X-ray, optical, radio and sub-TeV emission. Likewise, an emission of $(1.7\pm 1.0)\times 10^{53}$ erg originates from the accretion of the SN ejecta onto the companion, fast-spinning magnetized NS, creates the two events of ultra-relativistic jetted emissions (UPE I and UPE II), and the jetted GeV emission; see details in Table~\ref{tab:Summary}.

Therefore, BdHNe I are characterized by seven Episodes and four long-lasting emissions in the radio, optical, X-ray, and GeV bands. The total energy originating into the gravitational collapse of the CO$_{\rm core}$ in GRB 180720B is therefore of $(8.0\pm 1.0) \times 10^{53}$ erg, which classifies this source as a Giganova.

%%%%%%%%%%%%%%%%%%%%%%%%%%%%%%%%%%%%%%%%%%%%%%%%%%%
%\section*{Acknowledgements}
%We acknowledge ...
%%%%%%%%%%%%%%%%%%%%%%%%%%%%%%%%%%%%%%%%%%%%%%%%%%%
%\section*{Data Availability}

%\bibliographystyle{aasjournal}
%\bibliography{biblio}

\end{document}